\documentclass[a4paper,12pt]{article}
\usepackage{amssymb,rotating,graphics,epsfig,float,graphics}
\usepackage{amsmath, verbatim,a4wide}
\usepackage{imakeidx}
\title{Evolution to mirror-symmetric galaxies}
\author{Ferdinand Verhulst\\
Mathematisch Instituut, University of Utrecht\\PO Box 80.010, 3508 TA Utrecht, The Netherlands\\
email: f.verhulst@uu.nl}

\begin{document}

\maketitle
\abstract{ 
The evolution of a rotating axisymmetric galaxy from an asymmetric state to a state 
of mirror symmetry with respect to the galactic plane has as basic result that in the asymmetric 
initial state the 
perpendicular $z$ normal mode is  unstable for the $1:1$ and $1:2$ resonances. 
Dynamically this results in a transfer of mass and momenta towards the galactic plane. 
The timescale of evolution to symmetric equilibrium will determine in these cases the 
final distribution function describing position and velocities. In the case of the $1:1$ resonance we have in the final stage apart from the angular momentum integral 2 adiabatic invariants describing nonlinear 
dynamics. For the $1:2$ resonance the dynamics in the final stage is simpler,  apart from the angular 
momentum integral the dynamics is governed by the 2 actions. In the first sections the results of 
mathematical analysis have been summarised, examples of evolution are given in section 3.}\\


MSC classes:	34C14, 34C29, 34E10, 37J25, 37J65, 70H11, 85A05

\section{Modeling complex evolution} \label{sec1} 
The dynamics of galaxies cannot be understood without considering their evolution. 
There are many examples in astrophysics where evolution has produced systems with certain symmetries, 
 think of globular clusters, disk galaxies and elliptical galaxies. In the solar 
system planets usually take a spherical shape with rotation producing a certain flattening at the poles. 
Orbits of satellites around planets with tidal friction often tend to evolve to circular orbits in a plane while  
locking into $1:1$ resonance; this resonance is present for the Earth-Moon system where one rotation of the Moon around its axis corresponds with one revolution around Earth. \\ 
We consider the evolution of galaxies with, because of observations, special interest in the 
evolution of the position and velocity 
distributions of stars from a relatively asymmetric state to a symmetric one. \\
The evolution of galaxies is a very complex affair. {\em Our basic assumption is that the systems are already 
in an axisymmetric state of mass distribution and evolve slowly to mirror-symmetry with respect to 
the plane of the galaxy}. One can drop the assumption of axi-symmetry 
but this adds one degree of freedom and one timescale to the system with more computational possibilities. 
Note that different timescales are involved. 
For instance in the orbital evolution of a satellite around a heavy mass with tidal friction one can reach
 the state of $1:1$ resonance with the motion still far from planar. \\
Apart from the basic assumption our formulation of the Hamiltonian is completely 
general except for the choice of  explicit time-dependence.  
Changing the shape of time-dependence there is no difficulty in repeating the calculations.  \\
 
  The collisionless Boltzmann equation 
 describes  the evolution of the distribution of particles $f(t, {\bf x}, {\bf v})$ where ${\bf x}$ indicates the position, ${\bf v}$ the velocity; the collective 
 gravitational potential ruling the dynamics of the system is $\Phi(t, {\bf x})$, see \cite{BT} ch.~4.
With Lagrangian derivative $d/dt$ the equation is $df/dt =0$.  This is a first order 
partial differential equation with characteristics given by the Hamiltonian equations of motion. 
 The axisymmetry implies when introducing cylindrical coordinates $R, \phi, z$: 
 \begin{equation} \label{axi}
 \frac{\partial \Phi}{\partial \phi}  =0,
 \end{equation}
 producing the 
angular momentum integral $J$ (in \cite{BT} called $L_z$) enabling us to reduce the spatial 3-dimensional motion to 2 degrees of  
freedom (dof). The angular momentum integral  is: 
\begin{equation} \label{mom}
R^2 \dot{\phi} = J. 
\end{equation}
We want to describe the dynamical consequences of evolution to a 
 mirror-symmetric state by expanding around the 
circular orbits in the galactic plane at position $(R, z)= (R_0, 0)$ putting $R=R_0 + x$. 
To study evolution to symmetry we 
consider as a model the time-dependent two dof Hamiltonian: 
\begin{equation} \label{Ham} 
H = \frac{1}{2}(\dot{x}^2 +x^2) + \frac{1}{2}(\dot{z}^2 +\omega^2 z^2) -(\frac{1}{3}a_1x^3 +a_2xz^2) 
- e^{- \varepsilon^n t} (\frac{1}{3}a_3z^3 + a_4x^2z). 
\end{equation} 
The parameter $\varepsilon$ is small and positive, with $n$ it determines the timescale of evolution. 
We will choose $n=1$ or 2. 
The epicyclic frequency in the galactic plane has been scaled to 1, the vertical frequency is $\omega$, 
so $x$ corresponds with deviations of $R_0$, the radius of circular orbits. 
 The two velocity dispersions differ considerably; for velocity dispersions in the plane of the galaxy 
 see \cite{DMS06},  
 for dispersions in the halo \cite{B10}. Velocity dispersions in galaxies is an ongoing research topic.
To make the local analysis more transparent we rescale the coordinates $x = \varepsilon \bar{x}$ etc. Dividing by 
$\varepsilon^2$ and leaving out the bars we obtain the equations of motion: 
\begin{eqnarray} \label{eqs}
\begin{cases}
\ddot{x} + x & =   \varepsilon (a_1x^2 +a_2z^2) + \varepsilon   e^{- \varepsilon^n t} 2a_4xz,\\ 
\ddot{z} + \omega^2 z & =   \varepsilon 2a_2xz + \varepsilon  e^{- \varepsilon^n t} (a_3z^2 +a_4x^2).
\end{cases} 
\end{eqnarray} 
A more general formulation is given in \cite{V21} also including computational details.. For the mathematical  analysis one uses first and second order averaging, see \cite{SVM} or \cite{V23}. 
 
. If  we choose $a_1=1, a_2=-1, a_3=a_4=0$ we have the famous H\'enon-Heiles problem \cite{HH64}. 
The possible values of $\omega$ depend on the galactic potential 
constructed. An example describing an axisymmetric rotating oblate galaxy can be found in \cite{BT}
 eq. (3-50). \\ 
A time-independent  example with in the centre of the galaxy a very massive nucleus  
produces with mass $M$ large the potential: 
\begin{equation} \label{Mcentre}
\Phi  = \Phi_0 (\sqrt{R^2+z^2})+ \Phi_1(R, z), \Phi_0= - \frac{M}{\sqrt{R^2+z^2}},  
\end{equation} 
where, at least near the centre, $\Phi_1$ is small with respect to $\Phi_0$.  It is well-known that, 
assuming rotation and expanding in a 
neighbourhood of the centre and near the 
circular orbits we find for the epecyclic orbits and the orbits perpendicular to the galactic plane the $1:1$ resonance.  Outside the disk of the galaxy extending the galactic plane this resonance may again be prominent.

\section{Evolution of the dynamics to mirror-symmetry} \label{sec2} 
  Assuming $\omega \geq 1$ the resonances with most dynamical consequences are generally 
  $1:1$, $1:2$ and $1:3$, for the background of nonlinear resonance see \cite{SVM}, ch.~10. \\
It will be useful to introduce  polar coordinates 
\begin{equation} 
x= r_1 \cos(t+\psi_1),\, z= r_2 \cos(\omega t + \psi_2),
\end{equation}
and the actions $E_x, E_z$ by: 
\begin{equation} \label{actions}
E_x= \frac{1}{2}(\dot{x}^2 + x^2)= \frac{1}{2}r_1^2,\,E_z= \frac{1}{2}(\dot{z}^2 + \omega^2 z^2)= 
\frac{\omega^2}{2}r_2^2. 
\end{equation}
 The dynamics of prominent resonances produces periodic solutions, adiabatic invariants 
 and interesting stability changes during evolution. 
 
\subsection{The $1:1$ resonance} 
If the epicyclic frequency and the vertical frequency are equal or close, the $1:1$ resonance becomes important. 
The dynamics of the $1:1$ resonance turns out to be the most complicated case. 
 Depending on the system parameters  we find  
 normal modes (families of periodic solutions in the galactic plane and perpendicular to it) and stability 
 changes during evolution. 
Averaging the equations of motion we find that to $O(\varepsilon)$ all averaged terms vanish. 
Significant dynamics takes place on a longer timescale so we choose 
$n \geq 2$  to consider longer timescales, we have to use second order averaging. 
The combination angle $\chi = \psi_1 - \psi_2$ plays a part at second order. 
 Both time-dependent and  time-independent terms are strongly active on 
intervals of time $O(1/ \varepsilon^2)$. 
We summarise  from \cite{V79} the results for the end-stage of mirror-symmetry $a_3=a_4=0$. 
\begin{itemize} 
\item The equations of motion have in the mirror-symmetric case 2 adiabatic invariants: 
\begin{equation} \label{11Int1}
E_0= \frac{1}{2}(r_1^2 +r_2^2) = \frac{1}{2}(\dot{x}^2 +x^2 + \dot{z}^2 +z^2)
\end{equation} 
and: 
\begin{equation} \label{11Int2} 
I_3= r_1^2r_2^2 \cos 2 \chi + \alpha r_1^4 + \beta r_1^2, 
\end{equation}
with $\alpha, \beta$ rational functions of $a_1, a_2$. 
\item The epicyclic $x, \dot{x}$ normal mode is in the mirror-symmetric stage an exact solution, it is unstable for $- 1/3 < a_1/(3a_2) < 2/15$ and 
$1/3<  a_1/(3a_2) < 2/3$.
\item The vertical $z, \dot{z}$ normal mode is obtained as $O(\varepsilon)$ close to perpendicular motion. 
It is unstable for   $- 1/3 < a_1/(3a_2) < 1/3$. 
\item There are in-phase periodic solutions $\chi = 0, \pi$ existing and out-phase periodic solutions 
$\chi = \pi/2, 3 \pi/2$. 
\end{itemize} 
We expect the dynamics of the mirror-symmetric case to describe the orbits of system \eqref{eqs} on 
intervals of time larger 
than $1/ \varepsilon^2$. During evolution the in-phase and out-phase solutions are present with 
constant amplitude and are 
approaching periodicity. The dynamics will after a long time be governed by the angular momentum
integral \eqref{mom} and the 2 adiabatic invariants \eqref{11Int1}-\eqref{11Int2} producing 
a complicated velocity distribution and varying actions as demonstrated in the examples. 

\subsection{The $1:2$ resonance}

A prominent case for general 2 dof systems is the $1:2$ resonance ($\omega =2$).  
We find the active combination angles $\chi_1 = 2 \psi_1 - \psi_2$ and in the  
mirror-symmetric state $\chi_2 = 4 \psi_1 -2 \psi_2$.
First order averaging produces instability of the vertical $z$ normal mode ($x= \dot{x}=0$); 
the epicyclic $x$ normal mode does not exist on the 
interval of time $O(1/ \varepsilon)$ but emerges on a longer timescale.  
There are 2 adiabatic invariants valid on intervals $O(1/ \varepsilon)$: 
\begin{equation} \label{intave1}
\frac{1}{2}r_1^2 + 2r_2^2= E_0, \hspace{0.5cm} a_4r_1^2r_2 \cos \chi_1 = I_3,  
\end{equation} 
with constants $E_0, I_3$ and $r_1, r_2$ radii in polar coordinates, the asymmetry ($a_4$) is prominent. 
In the original coordinates we have: 
\[ \frac{1}{2}(\dot{x}^2 +x^2) + \frac{1}{2}(\dot{z}^2 + 4z^2)=E_0,\,\,  
a_4(x^2z- \dot{x}^2z + 2x\dot{x}\dot{z})=I_3. \] 
 On longer intervals of time the terms with 
coefficients $a_3, a_4$ will vanish and the symmetric terms become dominant. 
The first order averaged system admits families of solutions with constant amplitude and decreasing phases 
on intervals $O(1/ \varepsilon)$ if: 
\begin{equation} \label{resman1}
 \chi_1 = 0, \pi, r_1^2 = \frac{4}{3}E_0, r_2^2= \frac{1}{6}E_0. 
 \end{equation}
 The solutions with $\chi_1 =0$ are again called in-phase, the solutions with $\chi _1= \pi$ out-phase. 
 Second order averaging is needed to consider longer intervals of time and new phenomena. 
We summarise some results of \cite{V21}.
\begin{enumerate}

\item During an interval of time of order $1/ \varepsilon$ the integrals (adiabatic invariants) \eqref{intave1} are active,  
the system is dominated by the asymmetric $a_4$ term. On this interval of time it 
will govern the orbital dynamics and accordingly the corresponding distribution function in phase-space. 

\item On time intervals of order $1/ \varepsilon^2$ the time-independent 
system involving the coefficients $a_1, a_2$ dominates the dynamics. In \cite{TV00} it is shown that for 
this system, depending on $a_1, a_2$, 2 resonance manifolds can exist on the energy manifold 
In this case the resonance manifold 
with $4 \psi_1 - 2 \psi_2 =0$ has stable $2:4$ resonant periodic orbits surrounded by tori, 
for $4 \psi_1 - 2 \psi_2 = \pi$ the $2:4$ resonant 
periodic orbits also exist but are unstable. The resonance manifolds have size $O(\varepsilon)$, the dynamics takes 
place on intervals of time of order $1/ \varepsilon^3$.  
Outside the resonance manifolds the dynamics is characterised by the adiabatic invariants $E_x, E_z$.

\item
 The instability of the $z$ normal mode persists at second order but leads to stability in 
the final stage of mirror symmetry.. The epicyclic $x, \dot{x}$ normal mode 
does not exist in the first stage (time order $1/ \varepsilon$) but the normal mode emerges in the final 
stage of mirror symmetry. 

\end{enumerate} 

\section{Examples of evolution} \label{sec3} 

Dynamically interesting are the cases where we start in the beginning of evolution near an unstable orbit, 
say a normal mode, and 
move to a different dynamics when reaching a state close to mirror-symmetry. This changes the velocity 
distribution drastically. Such evolution can happen in the 
case of $1:1$ and of $1: 2$ resonance. \\
As described above these cases are dynamically different. The $1:1$ resonance when reaching 
a near mirror-symmetric state 
will still show nonlinear interaction between the modes; the normal modes persist but depending on the 
parameters can be stable or unstable. In the case of the $1:2$ resonance near mirror-symmetry implies 
that the dynamics has at the end of evolution the character of a $2:4$ resonance with little interaction of 
the modes; the normal modes persist and are stable. During evolution of the $1:2$ resonance the 
velocity distribution may change drastically. \\
The figures in the next subsections shows clearly the different types of dynamics of the $1:1$ and $1:2$ 
resonances in the final stage of mirror-symmetry.

\subsection{Evolution of the $1:1$ resonance}

\begin{figure}[H]
\begin{center}
\resizebox{14cm}{!}{
\includegraphics[width=7cm]{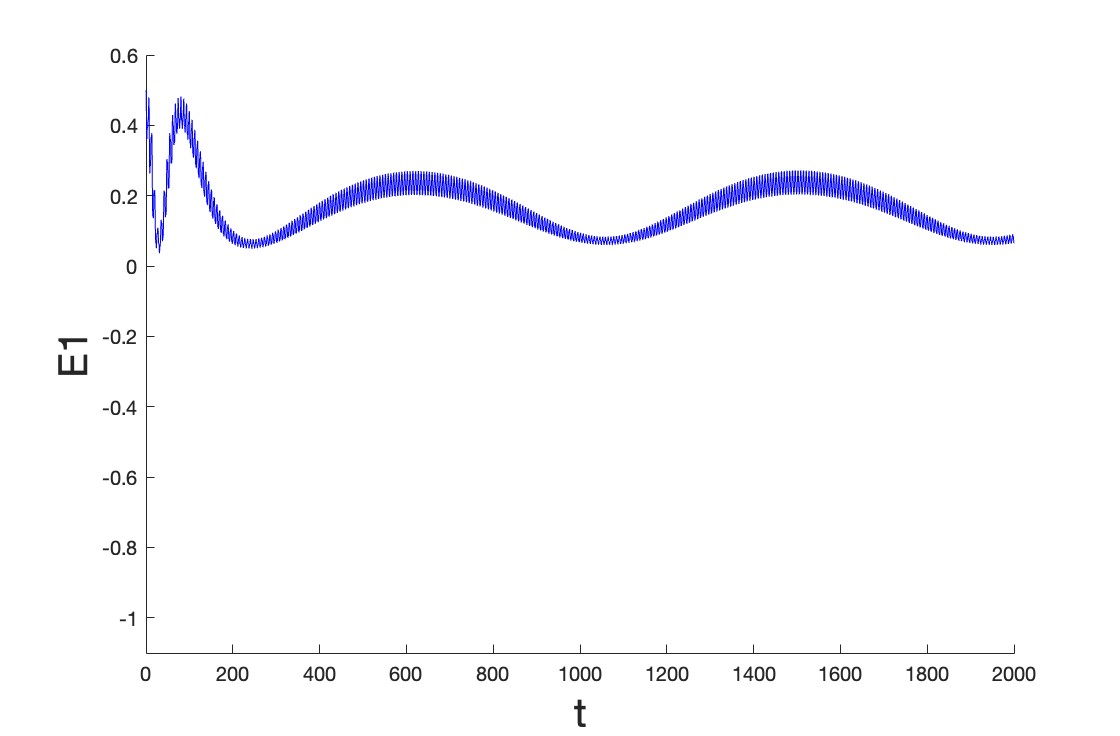} \, \includegraphics[width=7cm]{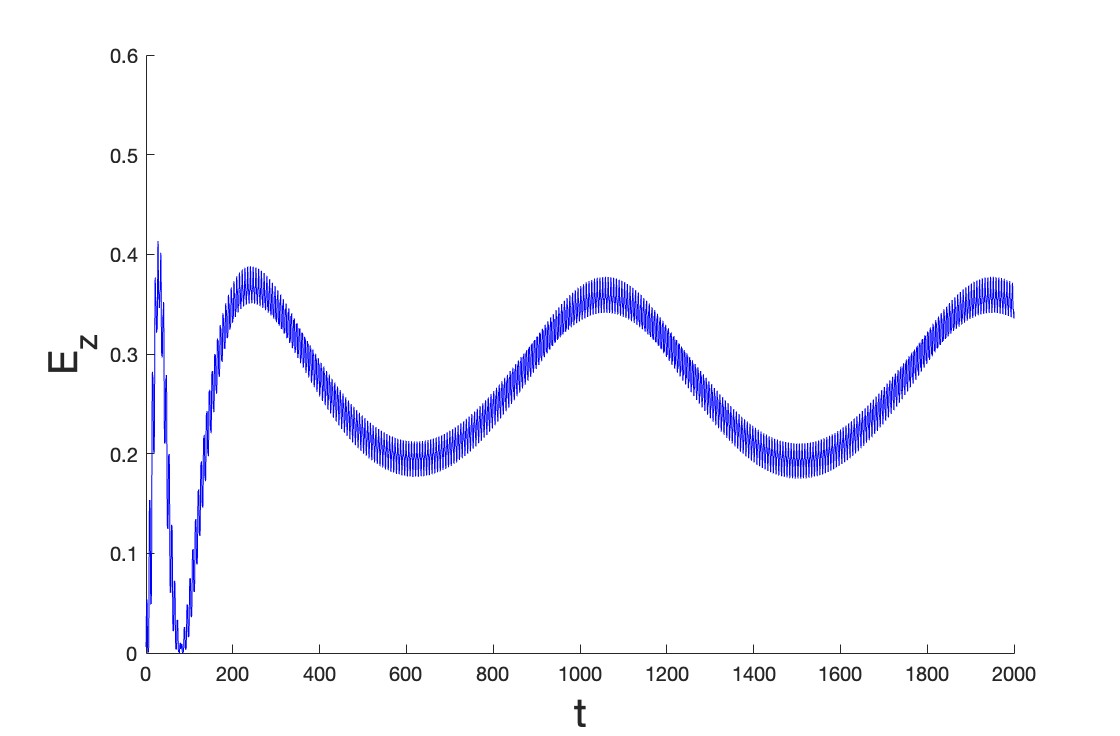}}
\caption{Left the behaviour of the action $E_x(t)$ of system \eqref{eqs} in $1:1$ resonance for the case $n=2$ near the epicyclic normal mode; 
  initial conditions $x(0)=1, z(0)=0.1, E_x(0)= 0.5, E_z(0)=0.005$. Parameter values $a_1=-1.5, a_2=1, 
  a_3=0, a_4=4, \varepsilon=0.1$. 
  Right we display the corresponding $z$ behaviour by plotting $E_z(t)$.
It takes around 200  timesteps to settle in the stationary state which is drastically different from the initial state. In the final stage the normal modes are stable.
 \label{fig1}}
\end{center}
\end{figure} 

\begin{figure}[H]
\begin{center}
\resizebox{14cm}{!}{
\includegraphics[width=7cm]{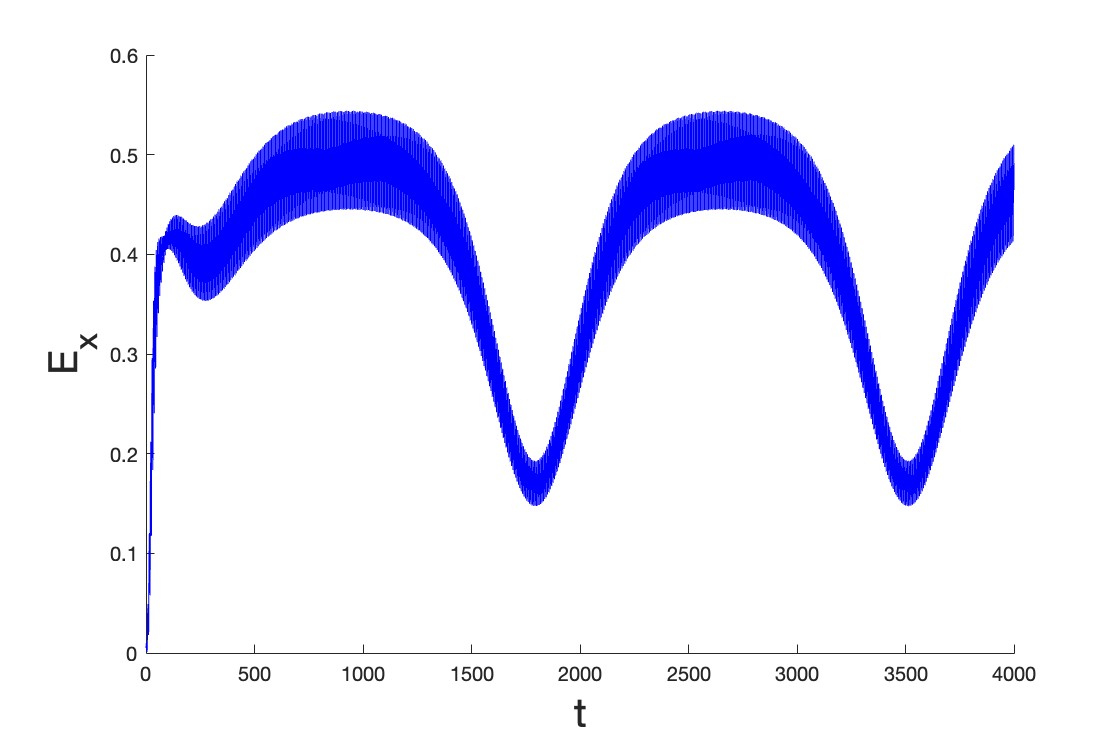} \, \includegraphics[width=7cm]{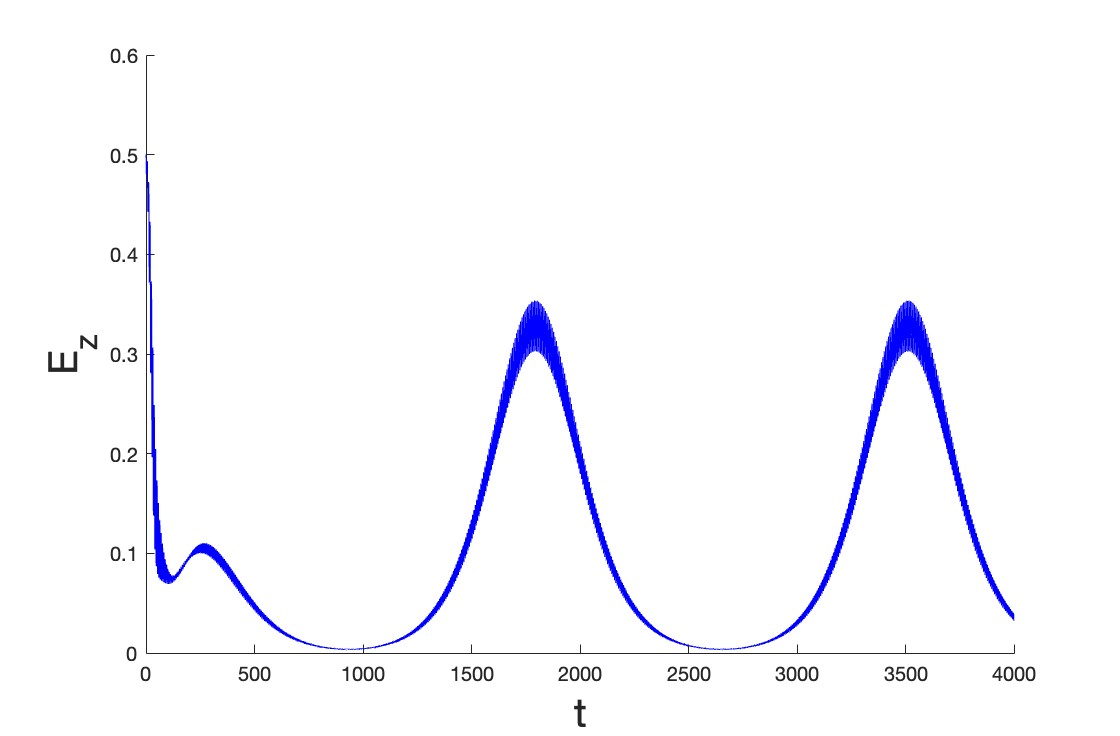}}
\caption{Left the behaviour of the action $E_x(t)$ of system \eqref{eqs} in $1:1$ resonance for the case $n=2$ starting near the vertical (z) normal mode; 
  initial conditions $x(0)=0.1, z(0)=1, E_x(0)= 0.005, E_z(0)=0.5$. Parameter values $a_1=-1.5, a_2=1, a_3=0, a_4=4, \varepsilon=0.1$. 
 In the final stage the normal modes are stable.
 \label{fig2}}
\end{center}
\end{figure}

In figs~\ref{fig1}-\ref{fig2} the choice of $a_1, a_2$ implies the system has stable epicyclic and stable vertical 
normal modes for the final mirror-symmetric case ($a_3=a_4=0$). 
Considering time evolution starting in an asymmetric state, our choice of $a_3, a_4$ keeps  the normal modes but 
destabilises them in the initial stage; we have  $a_1=-1.5, a_2=1, a_3=0, a_4=4$. When the 
time-dependent perturbation has become negligible 
the orbits have moved into general position. 
 The time-independent case will approximate the dynamics after a long initial interval of time, the in-phase 
 periodic solutions exist in this mirror-symmetric case and are stable. \\

\subsection*{Evolution to the H\'enon-Heiles dynamics} 

\begin{figure}[H]
\begin{center}
\resizebox{14cm}{!}{
\includegraphics[width=7cm]{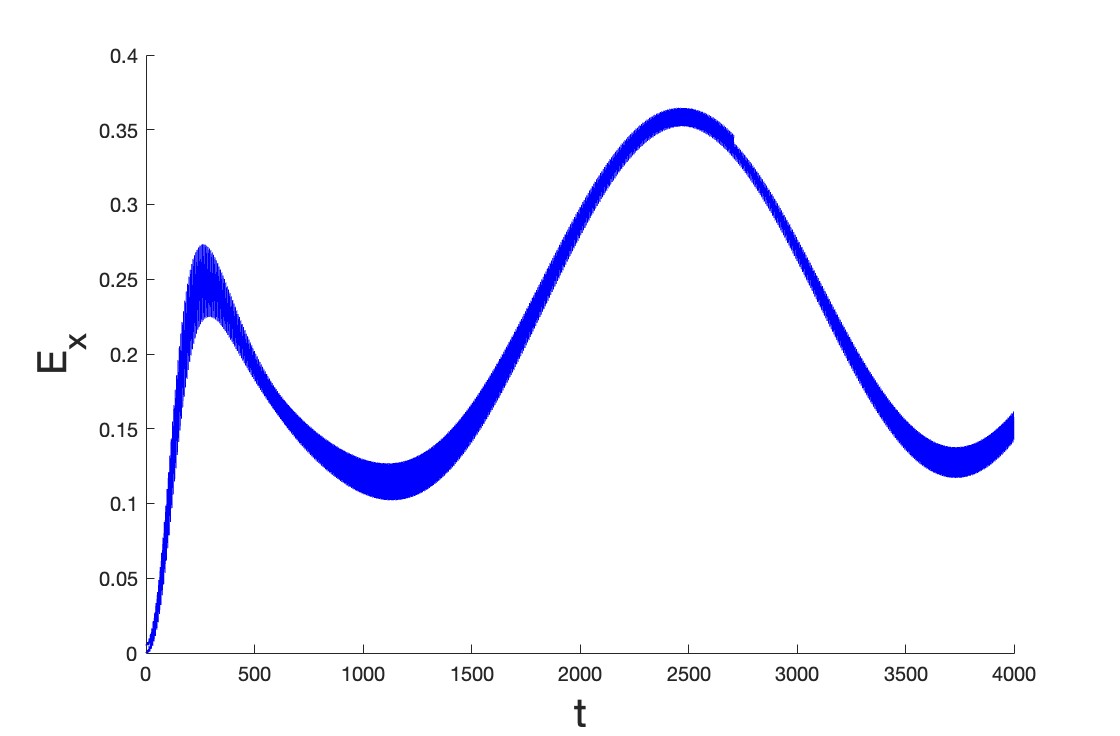} \, \includegraphics[width=7cm]{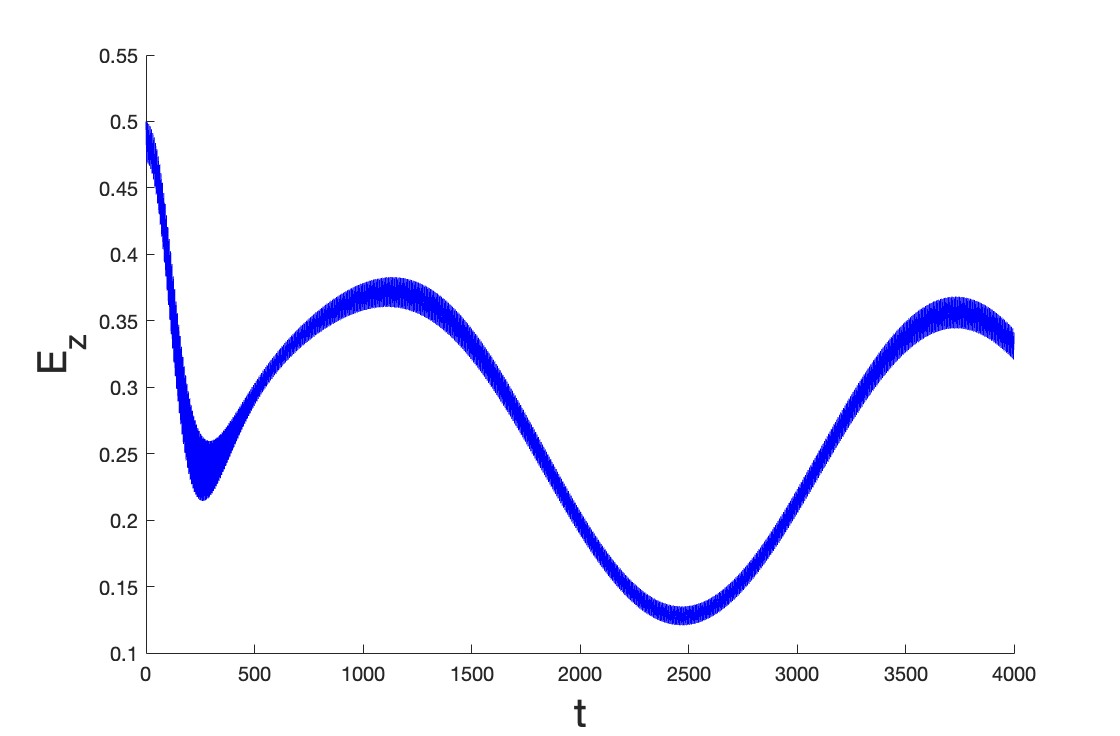}}
%\end{center}

\caption{Evolution to the H\'enon-Heiles system based on system \eqref{eqs} for the case 
$n=2, a_1=1, a_2=-1, a_3=1 , a_4=4 $ with 
$\varepsilon= 0.05$ intially close to motion perpendicular to the galactic plane  with 
$x(0)=0, z(0)=1$ and zero velocities. Left action $E_x$, right $E_z$. 
Because of the instability of the $z$  normal mode in the asymmetric initial stage the dynamics 
evolves to motion on a torus around a periodic solution in general position with in the final stage 
$E_x+E_z$ approximately constant. }  
 \label{HH}
\end{center}
\end{figure} 

The H\'enon-Heiles case is included as there are many details available 
in the literature on this system  \cite{HH64}.
If $a_3=a_4=0$ and $\varepsilon =1$ the system is chaotic; large-scale chaos sets in at 
$E_0 \geq 1/12= 0.083$, 
the energy 
manifold is bounded if $0 \leq E_0  \leq 1/6$. We consider the case of regular dynamics. 
In fig.~\ref{HH} (left) we present for $\varepsilon =0.05$ action $E_x(t)$ starting close to the unstable $z$ 
normal mode.  As we see in fig.~\ref{HH} (right) the motion into the halo becomes reduced whereas the component of motion in the galactic plane $E_x$ has become much larger.

\begin{figure}[H]
\begin{center}
\resizebox{14cm}{!}{
\includegraphics[width=7cm]{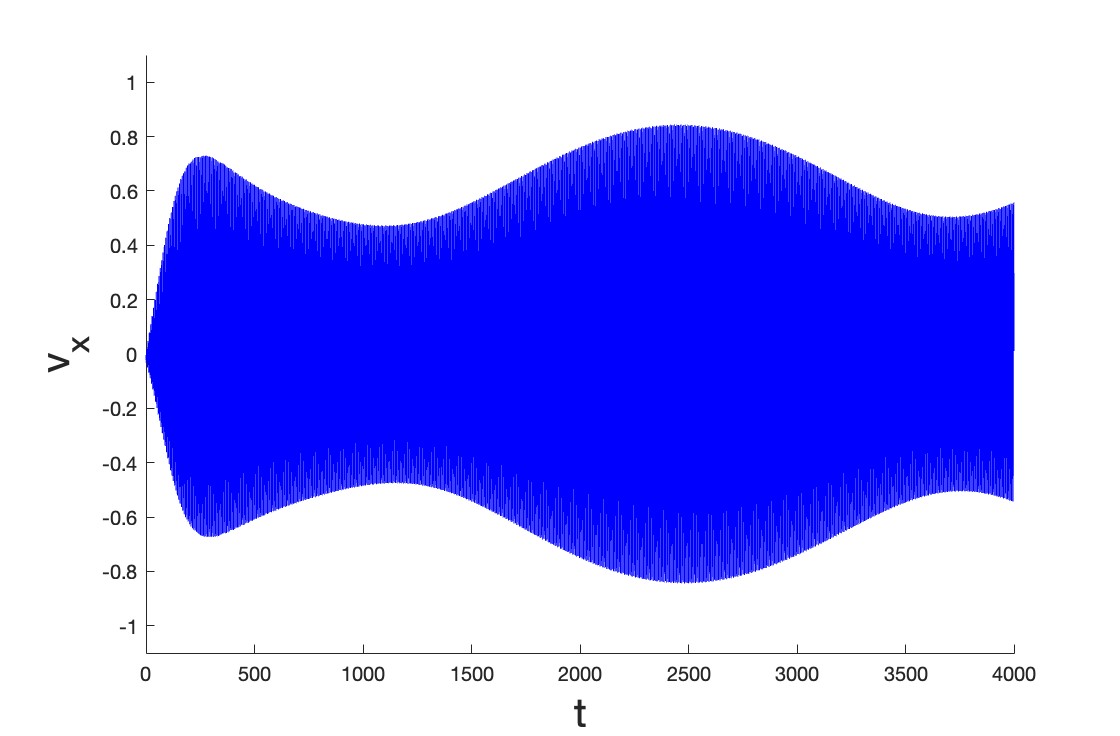} \, \includegraphics[width=7cm]{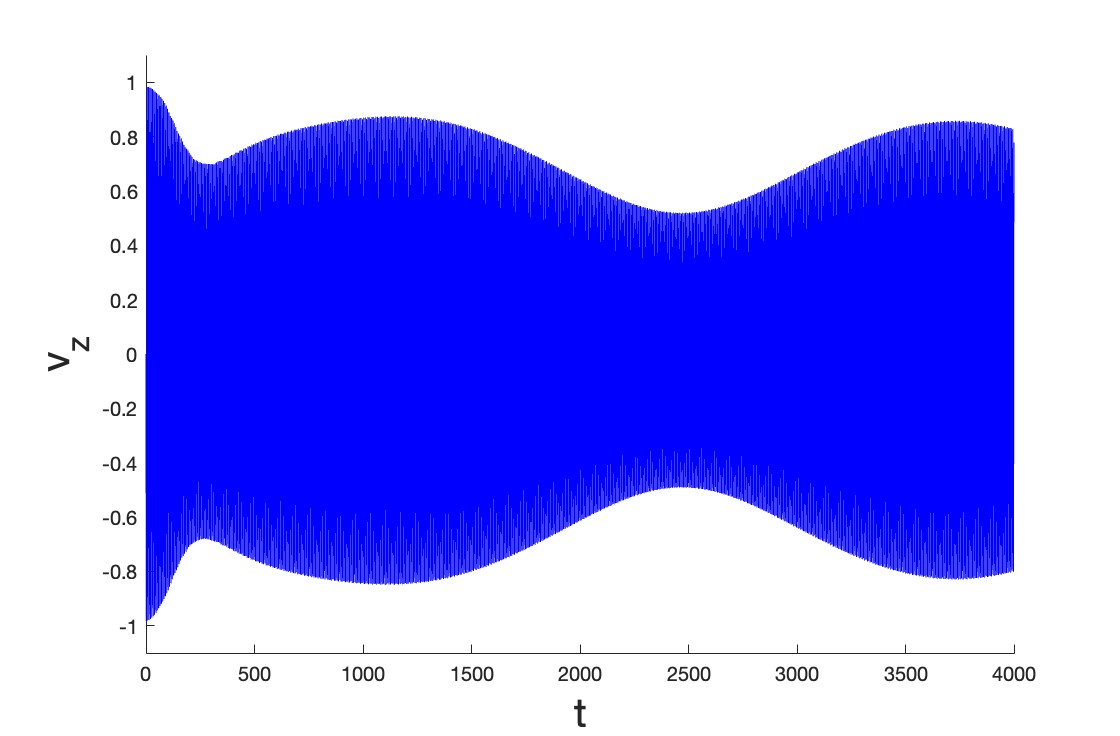}}
\caption{Evolution to the H\'enon-Heiles system of the velocity distribution based on system \eqref{eqs} 
with the same 
parameters as in fig.~\ref{HH}. The motion is initially close to motion perpendicular to the galactic plane  with 
$x(0)=0, z(0)=1$ and zero velocities. The velocity field corresponds with energy transfer to the galactic 
plane $x$ direction.}  
 \label{HHv}
\end{center}
\end{figure} 

The instability of the $z$ normal mode produces a drastic change of the velocity distribution. This is 
illustrated in fig.~\ref{HHv} where the evolution is shown of the velocities $v_x, v_z$ from the asymmetric 
case to the case of mirror-symmetry. 

Instead of the evolution to  the H\'enon-Heiles family one can formulate more general conclusions. In 
section \ref{sec2} we have listed open sets of parameters where the mirror-symmetric potentials 
$(a_3=a_4=0$) have existing normal $x$ and $z$ modes that are unstable. The examples 
given here are typical for the dynamics. 

\subsection{Evolution of the $1:2$ resonance} 
\begin{figure} 
%\widefigure
\begin{center}
\resizebox{14cm}{!}{
\includegraphics[width=7.5cm]{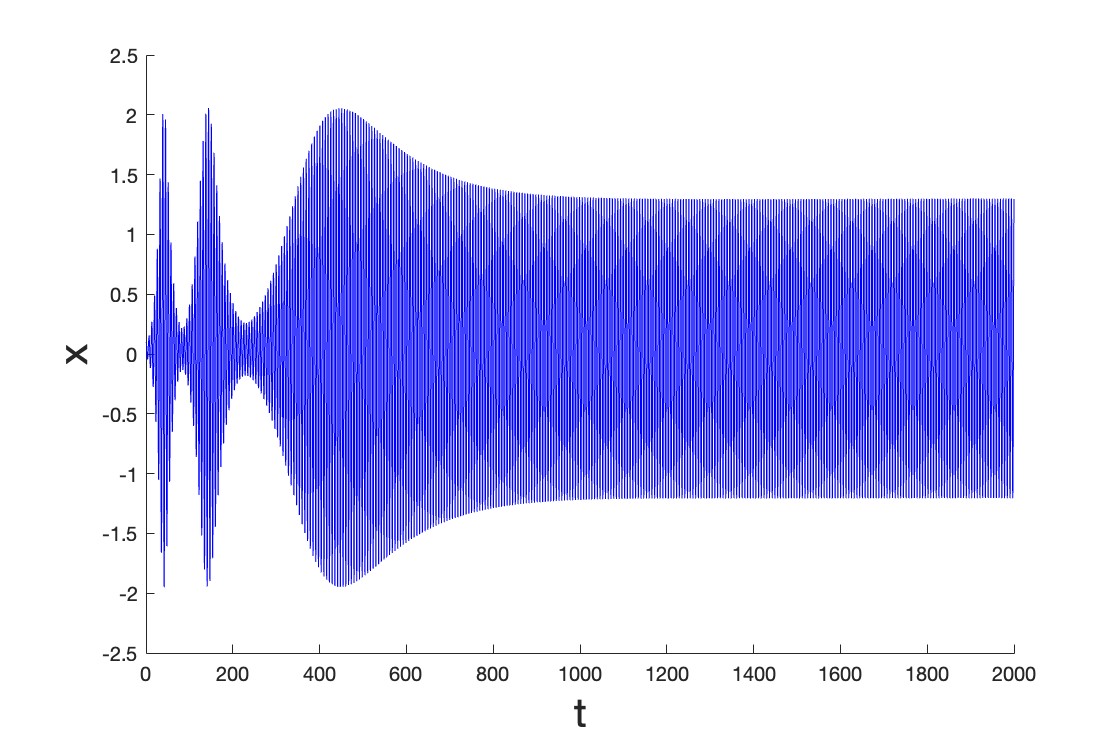} \,\includegraphics[width=7.5cm]{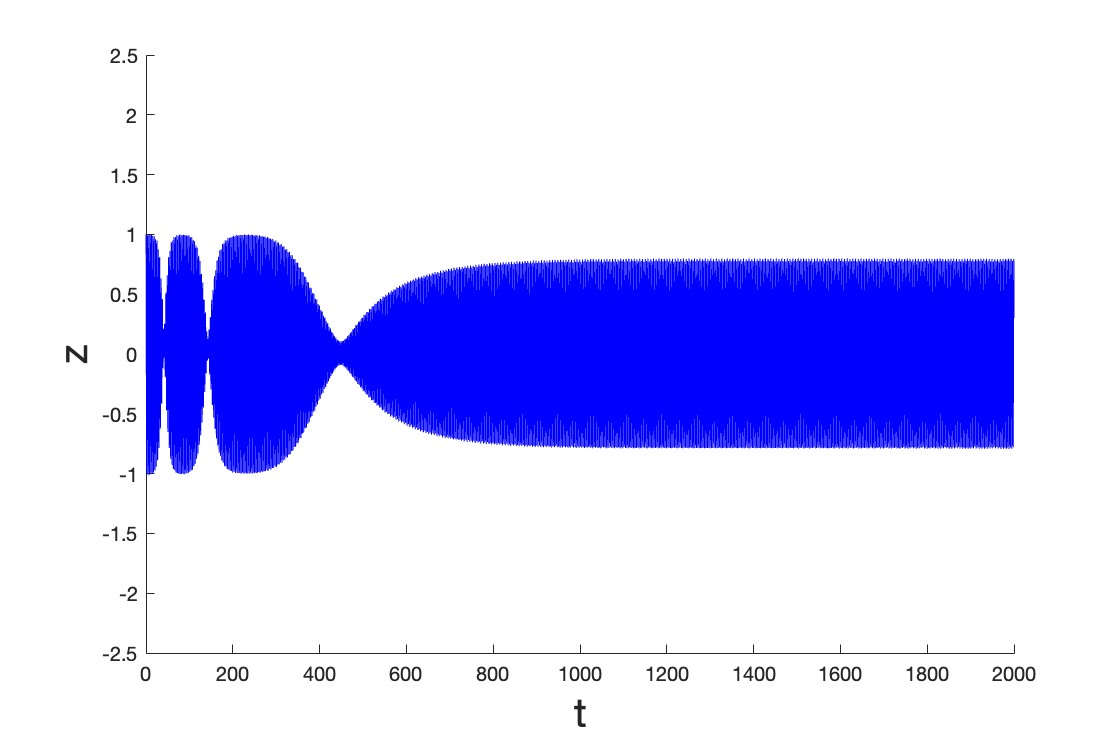} }
\caption{The $1:2$ resonance. The behaviour of system \eqref{eqs} starting  near the $z$-normal mode. 
We have $x(0)=0.1, z(0)=1$  with initial velocities zero, $ \varepsilon =0.08, \omega =2, 
a_1=0.5,  a_2= 1, a_3= -1, a_4=3$. 
The dynamics produces a position and velocity distribution that has changed considerably by the evolution 
 to mirror  symmetry.} \label{fig5}
\end{center} 
\end{figure}

.
On a timescale $O(1/ \varepsilon)$ the $z$ normal mode exists and is unstable; the epicyclic $x$ normal 
mode is not present but emerges during evolution. 
In fig.~\ref{fig5} we start near the $z$ normal mode position.
The solutions move into general position on tori around periodic solutions. The asymmetric 
past of the system has changed the overall dynamics drastically, but this depends strongly on the initial 
state where the potential is still asymmetric and the timescale of evolution.  \\
In fig.~\ref{fig6} we put $ \varepsilon = 0.05$ and $0.03$, the other parameters and initial 
conditions are as in fig.~\ref{fig5}. The evolution to 
mirror-symmetry takes longer as $\varepsilon$ is smaller than in fig.~\ref{fig5} with considerable influence 
on the position and 
velocity distribution. It is remarkable how sensitive the final dynamics is to the transition timescale of 
asymmetry to mirror-symmetry.

\begin{figure} \label{fig6}
\begin{center}
\resizebox{14cm}{!}{
\includegraphics[width=7cm]{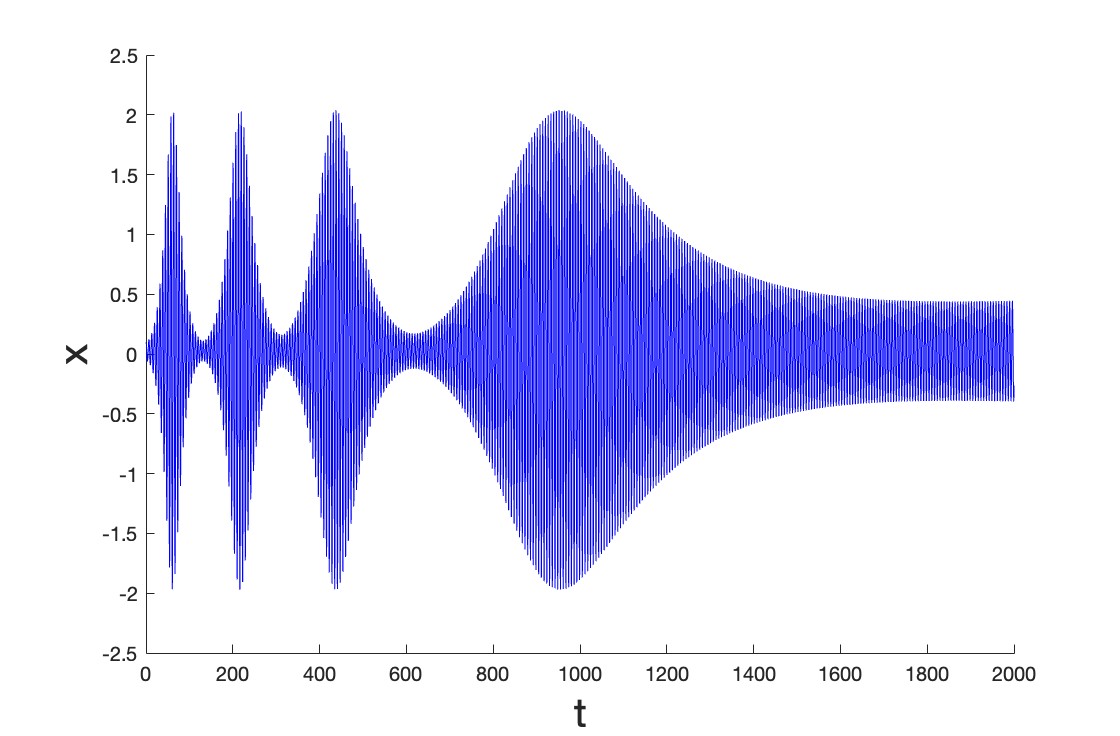} \, \includegraphics[width=7cm]{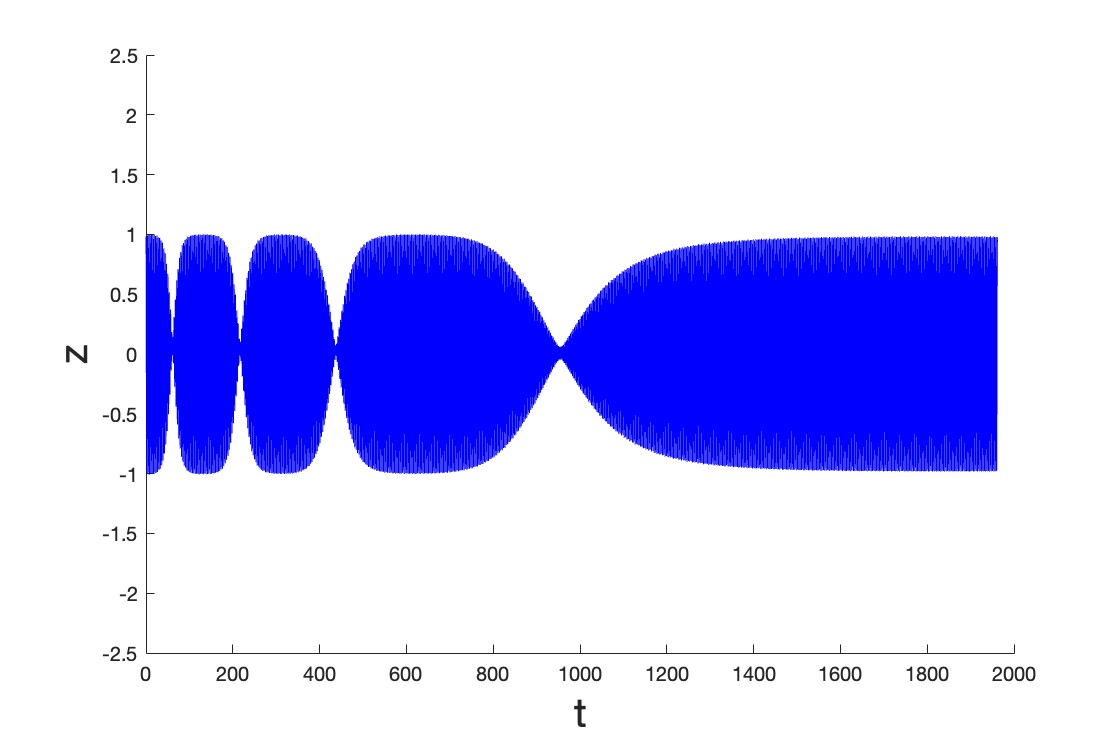}}\\
\resizebox{14cm}{!}{
\includegraphics[width=7cm]{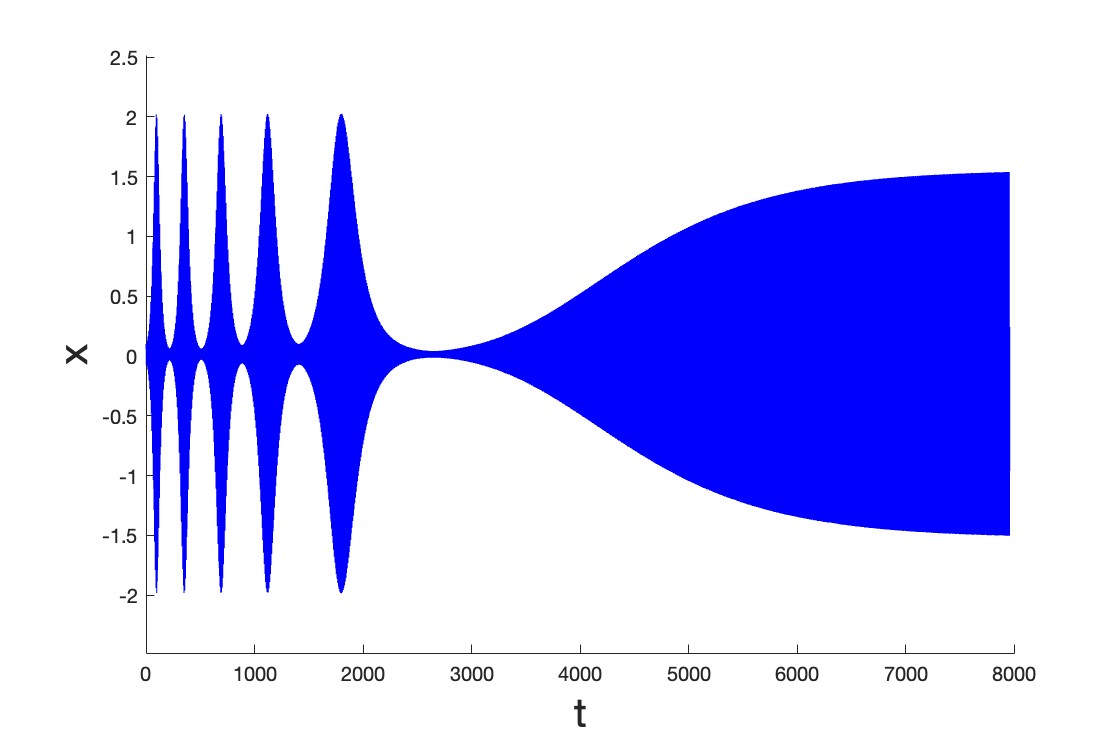} \, \includegraphics[width=7cm]{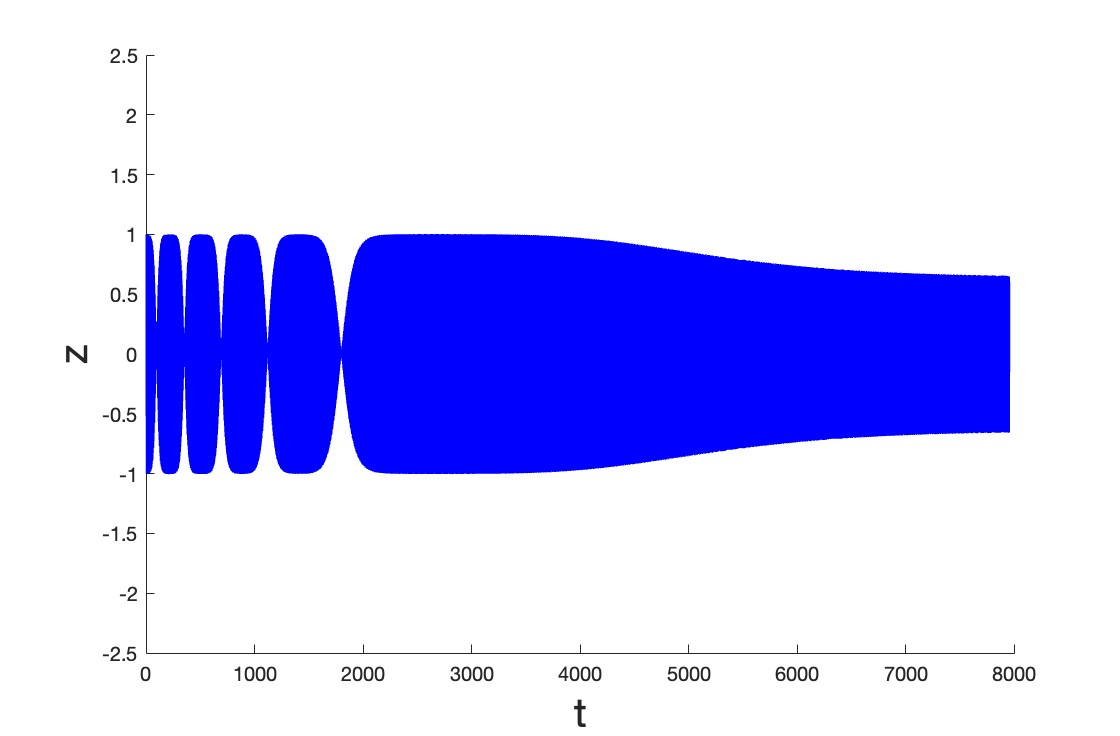}}

\caption{ The $1:2$ resonance.  
Decreasing $\varepsilon$ lengthens the transition time from asymmetry to symmetry 
and changes the transition dynamics. 
In fig.~\ref{fig5} we used $\varepsilon =0.08$. Keeping the other parameters and the initial conditions 
equal we show the 
$x(t), z(t)$ evolution for the cases $\varepsilon = 0.05$ (top), $0.03$ (below). } \label{fig6}
\end{center}
\end{figure}

\subsection{Evolution of the $1:3$ resonance}  
The $1:3$ resonance is dynamically different and less interesting. 
Most of the analysis can be deduced from  \cite{V79}, 
Putting $\omega =3$ in system \eqref{eqs}  we find after 2nd order averaging: 
\begin{equation} \label{ampl13} 
\dot{r}_1 = O(\varepsilon^3),\, \dot{r}_2 = O(\varepsilon^3). 
\end{equation} 
This remarkable result shows that for the $1:3$ resonance the slowly vanishing asymmetry of the 
potential {\em plays no 
part} for the amplitudes  to order $\varepsilon^3$. 
The theory of higher order resonance, see \cite{SVM} and  \cite{TV00}, shows that the 
combination angle $\chi_3= 6 \psi_1 -2 \psi_2$ plays a crucial role. 
 With high precision the 
distribution function at the $1:3$ resonance will depend on $J$ and the 2 actions.

\subsection{Consequences for the distribution function at the main resonances}  
We conclude and summarise. 
Suppose we start with a collection of particles (stars) characterised by a distribution function satisfying the 
Boltzmann equation that is collisionless  assuming negligible dynamical friction. The system is already 
in an axi-symmetric state but the evolution to mirror symmetry to the galactic plane is still going on. 
We have, apart from angular momentum $J$, two active integrals of motion strongly depending on the local 
resonance between epicyclic 
and vertical oscillations. The distribution function will be a function of the 3 integrals. Near resonance
the integrals will  change during evolution. 

A basic aspect of the evolution to mirror-symmetry is the instability near the $z$ 
normal mode (or in general, motion perpendicular to the galactic plane) in the original asymmetric state 
when close to $1:1$ or $1:2$ resonance. In this stage matter will be moved to the galactic disk and
 $x$-components of the velocities starting near perpendicular motion will grow. \\ 
 The timescale of evolution as given by the choice of $\varepsilon$ will also be very important for the 
 state of the final dynamics, see for examples figs \ref{fig5}-\ref{fig6}.

 Both the  $1:1$ or $1:2$ resonance show  this instability but there is also a difference.\\
 The $1:1$ resonance is characterised by nonlinear interaction of the 2 modes both in the asymmetric and in 
 the final symmetric state. In the final state we have for the 3 integrals of motion $J$, $E_0$ from 
 \eqref{11Int1} and $I_3$ from \eqref{11Int2}.\\  
 For the $1:2$ resonance we have in the initial asymmetric stage, (on a timescale of order 
 $1/ \varepsilon$) 
 the 3 integrals of motion $J, E_0, I_3$ with $E_0, I_3$ given by \eqref{intave1}. In the final stage, 
 say on timescales of order $1/ \varepsilon^3$, the integrals of motion will be $J, E_x, E_z$.  So 
 angular momentum and the 2 actions will dominate the dynamics of the $1:2$ resonance in the 
 mirror-symmetric stage.\\
 This also holds throughout for the evolution of the $1:3$ and other resonances from asymmetry to  
 mirror-symmetry. We leave out many interesting details of the higher order resonances, see \cite{V21} 
 for the theoretical background. The $O(\varepsilon)$ constancy of the actions in the $1:3$ case and in the 
 mirror-symmetric $1:2$ case can be illustrated numerically.

\medskip \noindent
{\bf Acknowledgement}\\

Funding:{This research received no external funding.} 

{The author declares no conflict of interest.}

\end{document}